\documentclass{aa}
\usepackage{graphicx}
\usepackage{amssymb}
\begin{document}
   \title{Planetary nebula or symbiotic Mira?\\Near infrared colours mark the difference\thanks{Based on observations collected at the European Southern Observatory, La Silla}}

   \author{S. Schmeja
          \and
          S. Kimeswenger}

   \offprints{S. Schmeja,\\ \email{stefan.j.schmeja@uibk.ac.at}}

   \institute{Institut f\"ur Astrophysik der Leopold-Franzens-Universit\"at Innsbruck, Technikerstr. 25, A-6020 Innsbruck, Austria (http://astro.uibk.ac.at)}

   \date{Received / accepted}

\abstract{ Nebulae around symbiotic Miras  look very much like genuine planetary nebulae, although they are formed in a slightly different way. We present near infrared photometry of known and suspected symbiotic nebulae obtained with the Deep Near Infrared Southern Sky Survey (DENIS). We demonstrate that the near infrared colours are an excellent tool to distinguish symbiotic from genuine planetary nebulae. In particular we find that the bipolar planetary nebulae M~2-9 and Mz~3 are in fact symbiotic Miras.
Further observations on prototype symbiotic Miras prove that the proposed classification scheme works generally.
\keywords{planetary nebulae: general  -- binaries: symbiotic -- stars: AGB and post-AGB -- stars: winds, outflows }
          }
\titlerunning{PN or symbiotic Mira?}
\maketitle


\section{Introduction}

\begin{figure}[t]
\resizebox{\hsize}{!}{\includegraphics{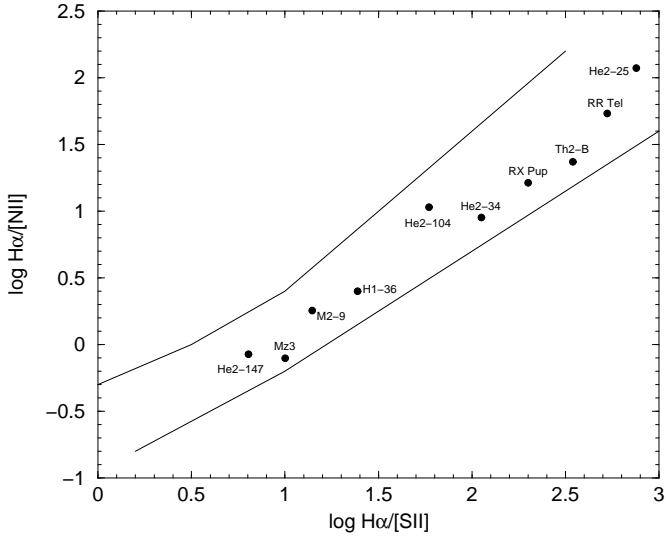}}
\caption{Diagnostic diagram for all objects from Fig.~\ref{2colour-diagram} with \mbox{(J--K)$_0$} $>$ 1\fm5. All the objects, including pure symbiotic Miras never classified as PNe, lie within the locus of genuine PNe (after Garc\'{\i}a Lario et al.\ \cite{Garcialario}). Spectral information is given in the literature.}
\label{diagdiag}
\end{figure}

In recent years, a number of nebulae around symbiotic Miras has been discovered, and several planetary nebulae (PNe) have been found to have a symbiotic binary nucleus (Corradi et al.\ \cite{Corradi99a}, \cite{Corradi00}, and references therein). 
Symbiotic Miras are interacting binary systems consisting of a cool Mira and a hot companion, in most cases a white dwarf. A small fraction (about 1\%) of the wind from the Mira is accreted by the compact companion. The vast majority of the Mira wind mass is ionized by the white dwarf and it forms a slowly expanding shell around the system. Many of those nebulae are extremely bipolar and have high velocity outflow features.
The nebulae around this class of symbiotic Miras are very similar to PNe in terms of 
morphology, excitation conditions, and chemical abundances, but they are no genuine PNe, since the gas is donated by a star still on the asymptotic giant branch and ionized by the hot companion whose PN is already dispersed in the interstellar matter. Due to many similarities nebulae around symbiotic Miras are hard to distinguish from classical PNe using optical spectroscopy. In a diagnostic diagram  used for gaseous nebulae (Garc\'{\i}a Lario et al.\ \cite{Garcialario}) plotting the intensity of the [\ion{N}{ii}] and [\ion{S}{ii}] lines relative to H$\alpha$,
all the symbiotic Miras from our sample lie exactly in the region of the PNe (Fig.~\ref{diagdiag}).
One big difference, however, are the near infrared (NIR) colours, as proposed by Whitelock \& Munari (\cite{WhiteMun}) using the JHK bands. In most cases direct evidence for a symbiotic binary central star still lacks: the cool (red) component is not visible, but detected by the modulation of the IR luminosity (e.\,g.\ He~2-104) or deduced from theoretical models (e.\,g.\ M~2-9). We finally are able to provide a tool to obtain the needed separation by using a longer wavelength base (as given by DENIS). As sample all planetary nebulae
with photo\-metry from the DENIS survey were taken. 
In addition, as ``prototype'' objects the known symbiotic 
Miras BI~Cru, RX~Pup, RR~Tel, \mbox{He~2-147}, and \mbox{H~1-36} 
have been observed using attenuating filters. 

\section{Observational Data} 
 DENIS (Deep Near Infrared Southern Sky Survey; Epchtein et al.\ \cite{Epchtein}) is a project which aims to survey the entire southern sky in the three near infrared bands Gunn-I (0.82~$\mu$m), J (1.25~$\mu$m), and K$\rm _s$ (2.15~$\mu$m). 
The observations are performed with the ESO 1\,m telescope at La Silla, Chile, which is fully dedicated to this project.\\
\begin{figure}[b]
\centerline{\includegraphics[width=6cm]{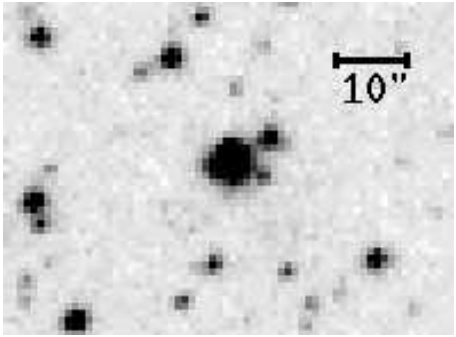}}
\caption{Contamination by background sources in aperture photometries using typical
diaphragm diameters of 20\arcsec\ in the NIR - example of He 2-147 (DENIS I band).}
\label{He2_147}
\end{figure}
In a work on NIR photometry of PNe we obtained a sample of 134 PNe observed with DENIS (Schmeja \& Kimeswenger \cite{Mexico}). The investigated PNe show a wide variety of morphologies, diameters, and distances, forming a pretty homogeneous sample. In any case, the flux from the whole nebula was measured. Photometry using the high resolution images allowed us to remove the background sources and thus to reduce the stellar background contamination to a minimum. This is a major improvement to the aperture photometries of PNe done so far. They often suffered from contamination by very red or strongly reddened background sources. Fig.~\ref{He2_147} shows \mbox{He~2-147} as an example how the field is contaminated by stars in normal aperture photometers.\\
For the calibration of our photometries we used the DENIS online zero points from La Silla, taking into account a small offset to the values derived at the Paris Data Analysis Centre (for details see Schmeja \& Kimeswenger \cite{Mexico}). 
The measured magnitudes were corrected for interstellar extinction using the extinction constants from the quick survey by Tylenda et al.\ (\cite{Tylenda}), except for He~2-25, Th~2-B, and 19W32, where we used the values given by Corradi (\cite{Corradi95}).\\
One of the main advantages of the DENIS instrument is the beam splitter, taking all three bands at the same time.
This makes the colours independent from minor errors due to changes of the atmospheric conditions. Moreover it gives us the colours of the objects independent from changes due to the pulsation of the Mira.
The instrument has fixed exposure times, thus it is restricted to a given magnitude range of 
13\fm8~$>$~K$\rm _s$~$>$~6\fm0 and 16\fm2~$>$~J~$>$~8\fm0.\\
Additional pointed observations of known symbiotic Miras have been obtained on June 5$^{\rm th}$, 7$^{\rm th}$, and 8$^{\rm th}$ 2001 with the same instrument using 5\fm2 attenuating filters. 
For the symbiotic Miras we used the interstellar extinction 
given by Whitelock (\cite{Wh87}).\\
The larger PNe sample is given in Schmeja \& Kimes\-wenger (\cite{Mexico}). 
The values of the 
attenuated observations, together with values from the literature (Munari et al.\ \cite{Munari}) are given in Table~\ref{photom}. As the objects are Miras, variations at different epochs are expected. 

\begin{table}[t]
\caption{Results of the additional photometries using the attenuator (see text).}
\label{photom}
{\small
\begin{tabular}{l|c c c|c c c}
Name & K$\rm _s$ & J-K$\rm _s$ & I-J & K & J-K & I-J \\
     &  \multicolumn{2}{c}{this work} &  & \multicolumn{3}{c}{Munari et al. (\cite{Munari})} \\
\hline
RX Pup & 3\fm05 & 2\fm96 & 3\fm56 & 3\fm13 & 3\fm16 & 3\fm49 \\
BI Cru & 4.\,\,53 & 2.\,\,36 & 1.\,\,87 & 4.\,\,88 & 2.\,\,57 & 1.\,\,54 \\
H 1-36 & 6.\,\,82 & 4.\,\,53 & 1.\,\,87 &  &  &  \\
RR Tel & 5.\,\,25 & 2.\,\,30 & 3.\,\,79 & 3.\,\,95 & 2.\,\,28 & 3.\,\,00 \\
He 2-147 & 4.\,\,58 & 2.\,\,46 & 4.\,\,64 & 5.\,\,18 & 2.\,\,41 & 4.\,\,18 \\
\end{tabular}
}
\end{table}
 
\section{Discussion}
 
\begin{figure*}[ht]
\sidecaption
\includegraphics[width=12cm]{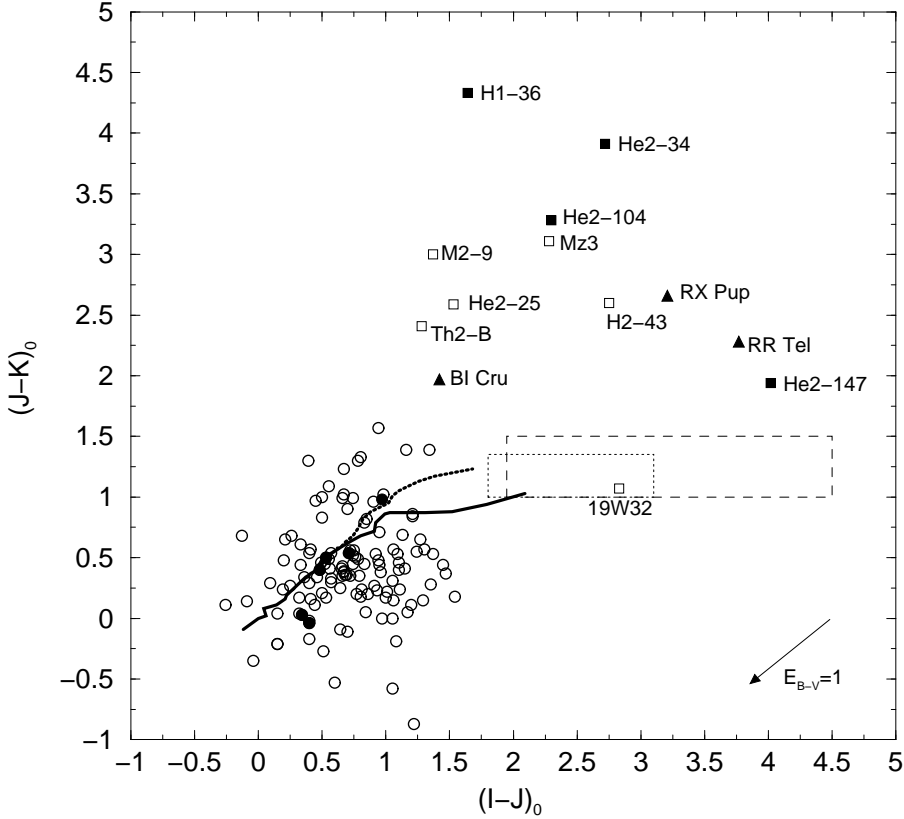}
\caption{~IJK two-colour diagram of genuine PNe (open circles), symbiotic Miras classified as PNe (squares), symbiotic Miras not classified as PNe (triangles), and suspected symbiotic Miras (open squares). Bipolar PNe that do not show any symbiotic behaviour are marked as filled circles.
Also shown are the positions of the stellar main sequence (solid line), the giants (dotted line), and the Miras and semiregular variables (dashed and dotted box, respectively; after Hron \& Kerschbaum \cite{HronKersch}). Those regions are defined not only by mean colours but consider the maximum boundaries of the tracks of Miras during the pulsation phases.}
\label{2colour-diagram}
\end{figure*}

Fig.~\ref{2colour-diagram} shows the dereddened NIR two-colour diagram of our sample of PNe. The photometric errors are about the size of the symbols. 
A vector corresponding to a reddening of $E_{\rm B-V}=1$ is also shown. It indicates the direction the objects would move in the diagram when increasing the extinction values.
The genuine PNe are located in a well defined region in the diagram, roughly in the range 
0\fm0~$\lesssim$~(I--J)$_0$~$\lesssim$~1\fm5 and  
-0\fm5~$\lesssim$~(J--K)$_0$~$\lesssim$~1\fm5. Especially, they are pretty well separated from the Miras, much better than in JHK two-colour diagrams used in the past. This is probably due to the longer wavelength base and the sensitivity of the I-band to non-thermal radiation. The known and suspected nebulae around symbiotic Miras, indicated with their names, however show peculiar colours both in (I--J)$_0$ and in (J--K)$_0$. They are in part quite far away from the bulk of other PNe, as well as from the classical Miras (except 19W32). So, the colours do not only represent the cool component. 
Whitelock (\cite{Wh87}) showed that ``classical'' symbiotic Miras have colours (J--K)~$>$~1\fm6.
This is also valid for those symbiotic Miras with PN-like nebulae in our sample.
The model of Whitelock (\cite{Wh87}) uses a 2500~K star and a 800~K enevlope with a circumstellar extinction starting at A$_{\rm K}\approx$~0\fm3 for RX~Pup and reaching up to 1\fm5. This simple extinction+cold emission model 
describes the 1-5\,$\mu$m colours well. Such a circumstellar extinction would lead to a circumstellar reddening of at least the same amount in case of (I--J). However, we find out colours that are even bluer than those of classical Miras, so a modification of this view is needed.
Also the spread in (I--J) is significantly higher than that in the pure NIR colour diagram.
The (I--J) is therefore giving the signature about the strength of the emission originating from
ionized gas and the hot component. Those parts cannot be obscured by the circumstellar shell as
proposed by the model of Whitelock. The more the object has an evolved nebula, the more
it moves to the upper left section of the diagram but not towards the normal PNe.
Thus it is also not just a stronger hot star+nebula. This is also valid for the objects 
H~1-36 and He~2-104 from the sample of ``classical'' symbiotic Miras (Whitelock \cite{Wh87}).
More sophisticated models are needed to describe the spectral energy distribution (SED).

\subsection{Symbiotic Miras also classified as PNe}

\object{He~2-104} (PN~G315.4+09.4), \object {He~2-147} (PN~G327.9-04.3), \object{H~1-36} (PN~G353.5-04.9), and \object {He~2-34} (PN~G274.1+02.5) are known symbiotic Miras also classified as PNe (Belczy\'{n}ski et al.\ \cite{Belczynski}; Acker et al.\ \cite{Acker}). \object{He~2-104}  has extremely bipolar lobes with high velocity outflows (Lutz et al.\ \cite{Lutz}; Corradi \& Schwarz \cite{CorrSchw93}; Corradi et al.\ \cite{Corradi00}). \object{He~2-147} is a symbiotic nova containing a Mira, the nebula around this object is a ring expanding with a velocity of $\sim$\,100~km\,s$^{-1}$ (Corradi et al.\ \cite{Corradi99b}, \cite{Corradi00}). \mbox{He~2-104} and H~1-36 are also the most extreme targets in 
Whitelock (\cite{Wh87}), where thus a circumstellar extinction of A$_{\rm V}\approx$10\fm0 is proposed.
H~1-36 has larger uncertainties than the other objects
(especially in the I band) since it required a complex background subtraction due to the bright nearby star \object{HD~161892}.

\subsection{Symbiotic Miras not classified as PNe}
\object{BI~Cru}, \object{RR~Tel}, and \object{RX~Pup} are typical symbiotic Miras, they have been observed in addition to our PNe sample. BI~Cru and RX~Pup have resolved nebulae (Corradi et al.\ \cite{Corradi99a}), but unlike the objects mentioned above, they have not been classified as PNe. BI~Cru is, due to the pulsation of the Mira, slightly brighter than in the measurements of Munari et al.\ (\cite{Munari}), while RR~Tel is significantly fainter, but the colours correspond quite well in both cases.

\subsection{PNe suspected to be symbiotic Miras}
\object{Mz~3} (PN~G331.7-01.0) shows morphological similarities with He~2-104 and M~2-9. It also has bipolar lobes with hypersonic outflows up to 500~km\,s$^{-1}$ (Redman et al.\ \cite{Redman}). It is suspected to have a symbiotic binary nucleus (Balick, private comm.). 
\\
The ``Butterfly'' nebula \object{M~2-9} (PN~G010.8+18.0) has two narrow bright lobes and has changed its shape drastically since its discovery in 1947 (Doyle et al.\ \cite{Doyle}). Balick (\cite{Balick}) identified its central source as a possible symbiotic star. In a recent model (Livio \& Soker \cite{LivioSoker}) the nucleus of M~2-9 is supposed to be a system of an AGB or post-AGB star and a hot white dwarf.\\
\object{H~2-43} (PN~G003.4-04.8) is classified as a suspected symbiotic star by 
Belczy\'{n}ski et al.\ (\cite{Belczynski}).
Our results show clearly that these three nebulae belong to the group of symbiotic Miras.\\

\object{He~2-25} (PN~G275.2-03.7), \object{Th~2-B} (PK~307-01.1), and \object{19W32} (PN~G359.2+01.2) are suspected to be symbiotics (Corradi \cite{Corradi95}), which is supported by our results. Corradi (\cite{Corradi95}) notes strong similarities in the spectra of He~2-25, Th~2-B, and M~2-9. In our diagram they also seem to form a group, as they are lying relatively close to each other. \object{19W32}, however, is placed among the Miras and red semiregular variables, clearly separated from the normal PNe as well as from the symbiotic Miras. The extinction towards this object is rather uncertain and might move it slightly in the two-colour diagram. In any case, 19W32 is not a genuine PN and may quite possibly belong to the class of symbiotic nebulae.
\\

\subsection{Other Bipolar Nebulae}

From our sample bipolar nebulae that do not show any symbiotic behaviour according to the two-colour diagram are \object{NGC~6537} (PN~G010.1+00.7), \object{Sa~2-237} (PN~G011.1+07.0), \object{M~1-57} (PN~G022.1-02.4), \object{NGC~2440} (PN~G234.8+02.4), \object{He~2-36} (PN~G279.6-03.1), and \object{MyCn~18} (PN~G307.5-04.9). They are indicated as filled circles in Fig.~\ref{2colour-diagram}. Thus the sometimes supposed conclusion that all bipolar PNe can be attributed to symbiotic stars is certainly wrong.\\
A special case is \object{SaSt~1-1} (PN~G249.0+06.9). It is a nebula around a symbiotic star (Schwarz \cite{Schwarz}; Corradi et al.\ \cite{Corradi99a}), but in our two-colour diagram it shows colours like a genuine PN. The cool component of this symbiotic system, however, is not as in the other cases a Mira, but a yellow star of G5 spectral type (M\"urset \& Schmid \cite{MuersetSchmid}). This may be the reason for the ``normal'' colours; it is consistent with the conclusions of Corradi et al.\ (\cite{Corradi99a}) that the nebulae around yellow symbiotics might well be the PNe of the presently hot white dwarfs of these systems.

\section{Conclusion}
We show that nebulae around symbiotic Miras have NIR colours completely different from genuine PNe. 
They are characterized in the two-colour diagram like ``normal'' symbiotic Miras, but well separated from classical Miras. 
Comparison of our measurements to those from the literature show that the magnitudes can differ significantly due to the Mira variability, whereas the colour indices are more or less the same.
Thus the colour is a unique idenitifier if obtained through simultaneous or quasi simultaneous measurements.
In contrast to classical Miras (Whitelock et al.\ \cite{Wh00}) there is no correlation of the NIR colours with the pulsation period.\\
We consider the NIR colours to be a unique tool to distinguish between symbiotic Miras and PNe.
In particular it turns out that M~2-9 and Mz~3, so far widely regarded as classical PNe, in fact belong to the symbiotic Miras. Nevertheless, further investigations of these objects, especially to detect the cool companions and to study the variations in the radial velocity, are needed. The models to describe the 1-5\,$\mu$m SED proposed so far do not work anymore for those objects considering the I band. Thus they are either a special class of symbiotic Miras, or the most extreme targets within a larger class. Ideas about long-term variations as discussed for \object{RR~Tel} and \object{IRAS~13568-6232} by Leeber et al.\ (\cite{Leeber}) have to be followed in detail, too.

\begin{acknowledgements}
The DENIS project is partly funded by the European Commission through {\it SCIENCE} and {\it Human Capital and Mobility} plans under grants CT920791 and CT940627. It is supported by INSU, MEN and CNRS in France, by the
State of Baden-W\"urttem\-berg in Germany, by DGICYT in Spain, by CNR in Italy, by the {\it Fonds zur F\"or\-de\-rung der wis\-sen\-schaft\-li\-chen For\-schung} and the {\it Bun\-des\-mini\-sterium f\"ur Bil\-dung, Wis\-sen\-schaft und Kultur} in Austria, by FAPESP in Brazil, by OTKA grants F-4239 and F-013990 in Hungary, and by the ESO C\&EE grant A-04-046.
\end{acknowledgements}


\begin{thebibliography}{}

\bibitem[1992]{Acker}
Acker, A., Ochsenbein, F., Stenholm, B., et al., 1992, Strasbourg-ESO Catalogue of Galactic Planetary Nebulae, ESO Garching 

\bibitem[1989]{Balick}
Balick, B., 1989, AJ, 97, 476

\bibitem[2000]{Belczynski}
Belczy\'{n}ski, K., Miko{\l}ajewska, J., Munari, U., Ivison, R.~J., \& Friedjung, M., 2000, A\&AS, 146, 407

\bibitem[1995]{Corradi95}
Corradi, R.~L.~M., 1995, MNRAS, 276, 521

\bibitem[1993]{CorrSchw93}
Corradi, R.~L.~M., \& Schwarz, H.~E., 1993, A\&A, 268, 714

\bibitem[1999a]{Corradi99a} 
Corradi, R.~L.~M., Brandi, E., Ferrer, O.~E., \& Schwarz, H.~E., 1999a, A\&A, 343, 841 

\bibitem[1999b]{Corradi99b}
Corradi, R.~L.~M., Ferrer, O.~E., Schwarz, H.~E., Brandi, E., \& Garc\'{\i}a, L., 1999b, A\&A, 348, 978

\bibitem[2000]{Corradi00} 
Corradi, R.~L.~M., Livio, M., Schwarz, H.~E., \& Munari, U., 2000, in Asymmetrical Planetary Nebulae II: From Origins to Microstructures, eds. J.H. Kastner, N. Soker, and S. Rappaport, ASP Conf.\ Ser., 199, 175

\bibitem[2000]{Doyle}
Doyle, S., Balick, B., Corradi, R.~L.~M., \& Schwarz, H.~E., 2000, AJ, 119, 1339

\bibitem[1997]{Epchtein}
Epchtein, N., de Batz, B., Capoani, L., et al., 1997, The Messenger, 87, 27

\bibitem[1991]{Garcialario}
Garc\'{\i}a Lario, P., Manchado, A., Riera, A., Mampaso, A., \& Pottasch, S.~R., 1991, A\&A, 249, 223

\bibitem[1994]{HronKersch}
Hron, J., \& Kerschbaum, F., 1994, Ap\&SS, 217, 137

\bibitem[1996]{Leeber}
 Leeber, D.~M., Whittet, D.~C.~B., Prusti, T.,
 Kilkenny, D., \& Whitelock, P.~A., 1996, ApJ, 463, L25

\bibitem[2001]{LivioSoker}
Livio, M., \& Soker, N., 2001, ApJ, 552, 685

\bibitem[1989]{Lutz}
Lutz, J.~H., Kaler, J.~B., Shaw, R.~A., Schwarz, H.~E., \& Aspin, C., 1989, PASP, 101, 966

\bibitem[1999]{MuersetSchmid}
M\"urset, U., \& Schmid, H.~M., 1999, A\&AS, 137, 473

\bibitem[1992]{Munari}
 Munari, U., Iudin, B.~F., Taranova, O.~G., et al., 1992, A\&AS, 93, 383

\bibitem[2000]{Redman}
Redman, M.~P., O'Connor, J.~A., Holloway, A.~J., Bryce, M., \& Meaburn, J., 2000, MNRAS, 312, L23

\bibitem[2001]{Mexico}
Schmeja, S., \& Kimeswenger, S., 2001, Rev.\ Mex.\ Astron.\ Astrofis.\ (Ser.\ de Conf.), in press

\bibitem[1991]{Schwarz}
Schwarz, H.~E., 1991, A\&A, 243, 469

\bibitem[1992]{Tylenda}
Tylenda, R., Acker, A., Stenholm, B., \& K\"oppen, J., 1992, A\&AS, 95, 337

\bibitem[1987]{Wh87}
Whitelock, P.~A., 1987, PASP, 99, 573 

\bibitem[1992]{WhiteMun}
Whitelock, P.~A., \& Munari, U., 1992, A\&A, 255, 171 

\bibitem[2000]{Wh00}
Whitelock, P.~A., Marang, F., \& Feast, M., 2000, MNRAS, 319, 728

\end{thebibliography}
\end{document}